\documentclass[manuscript]{acmart}

\AtBeginDocument{%
  \providecommand\BibTeX{{%
    \normalfont B\kern-0.5em{\scshape i\kern-0.25em b}\kern-0.8em\TeX}}}

\copyrightyear{2023}
\acmYear{2023}
\setcopyright{rightsretained}
\acmConference[TEI '23]{TEI '23: Proceedings of the Seventeenth International Conference on Tangible, Embedded, and Embodied Interaction}{February 26-March 1, 2023}{Warsaw, Poland}
\acmBooktitle{TEI '23: Proceedings of the Seventeenth International Conference on Tangible, Embedded, and Embodied Interaction (TEI '23), February 26-March 1, 2023, Warsaw, Poland}\acmDOI{10.1145/3569009.3573123}
\acmISBN{978-1-4503-9977-7/23/02}

\begin{document}

\title{PullupStructs: Digital Fabrication for Folding Structures via Pull-up Nets}


\author{Lauren Niu}
\authornote{Both authors contributed equally to this research.}
\email{lauren.l.niu@gmail.com}
\affiliation{%
  \institution{Harvard University}
  \city{Cambridge}
  \state{Massachusetts}
  \country{USA}
  \postcode{02138}
}

\author{Xinyi Yang}
\email{xinyiyang@gsd.harvard.edu}
\authornotemark[1]
\affiliation{%
  \institution{Harvard University}
  \city{Cambridge}
  \state{Massachusetts}
  \country{USA}
  \postcode{02138}
}

\author{Martin Nisser}
\affiliation{%
  \institution{Massachusetts Institute of Technology}
  \city{Cambridge}
  \state{Massachusetts}
  \country{USA}
  \postcode{02138}
}
 
\author{Stefanie Mueller}
\affiliation{%
  \institution{Massachusetts Institute of Technology}
  \city{Cambridge}
  \state{Massachusetts}
  \country{USA}
  \postcode{02138}
}

\renewcommand{\shortauthors}{Xinyi Yang, et al.}

\begin{abstract}
In this paper, we introduce a method to rapidly create 3D geometries by folding 2D sheets via pull-up nets. Given a 3D structure, we unfold its mesh into a planar 2D sheet using heuristic algorithms and populate these with cutlines and throughholes. We develop a web-based simulation tool that translates users' 3D meshes into manufacturable 2D sheets. After laser-cutting the sheet and feeding thread through these throughholes to form a pull-up net, pulling the thread will fold the sheet into the 3D structure using a single degree of freedom. We introduce the fabrication process and build a variety of prototypes demonstrating the method's ability to rapidly create a breadth of geometries suitable for low-fidelity prototyping that are both load-bearing and aesthetic across a range of scales. Future work will expand the breadth of geometries available and evaluate the ability of our prototypes to sustain structural loads.

\end{abstract}



\maketitle

\begin{figure}[H]
  \includegraphics[width=\textwidth]{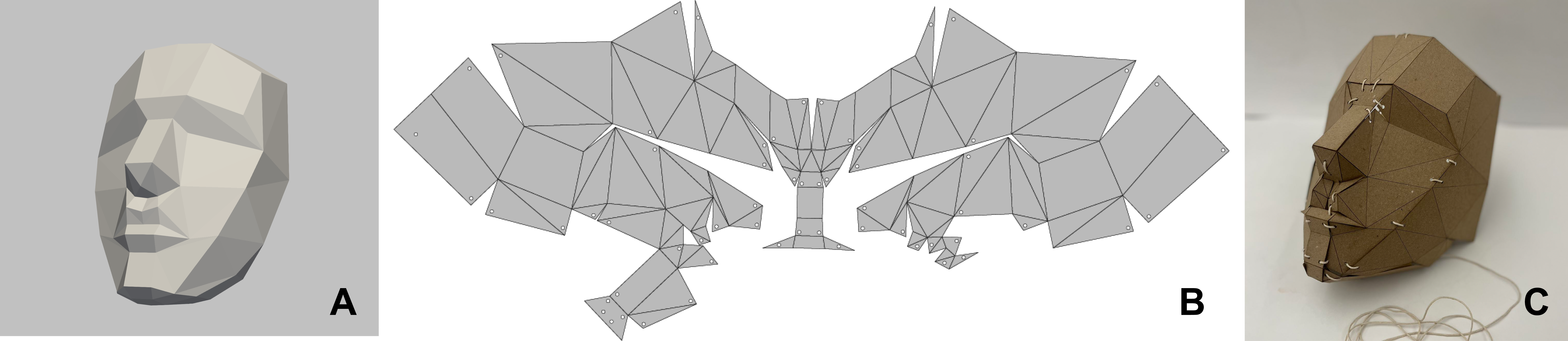}
  \centering
  \caption{Our digital fabrication pipeline enables users to rapidly prototype 3D objects. It features (A) a web-based tool for uploading and rendering 3D meshes. The tool executes our unpacking algorithm which unfolds the mesh into a (B) planar sheet partitioned into faces with through-holes in them. (C) We fabricate these unfolded geometries on a laser cutter, route thread through the holes, and pull the thread to fold the sheet into its target structure.}
  \label{fig: teaser}
\end{figure}

\section{Introduction}

Laser-cutting is one of the most commonly used manufacturing processes used in personal fabrication~\cite{baudisch2017personal}, and the fabrication speed and material choices available for laser-cutting make it a compelling substitute to 3D printing. However, unlike the latter, fabricating 3D structures from laser-cut parts may demand significant user intervention that is both time-consuming and technically cumbersome. 

A major push in HCI research has recently been made to address these bottlenecks in speed and domain knowledge to assembling laser-cut structures. One method of addressing these challenges simultaneously is by automating the folding procedures in the laser cutter itself, using the laser in an attenuated power setting to heat fold lines until they become compliant and the substrate folds into a target 2.5D geometry via gravity~\cite{mueller2013laserorigami,nisser2021laserfactory}. Similarly, another promising mechanism for improving the speed and ease of fabricating fully 3D structures involves laser-cutting joints and folds directly in place, permitting users to then fold the 2D laser-cut sheet into a target shape~\cite{abdullah2022hingecore,10.1145/3526113.3545695}. However, these methods have only been utilized for structures requiring a few dozen folds to date, and the nature of these processes to require users to manually fold each hinge individually may limit the efficacy and adoption of these processes for high-resolution structures. 

In this work, we introduce a method to rapidly create 3D geometries from 2D sheets using pull-up nets: a string routed through the planar faces which can be pulled by a user to fold the sheet into its target 3D structure. This provides a way to fold a sheet into its target shape using common string or nylon, using just \textit{a single actuated degree of freedom} controlled by a user. Researchers have also previously leveraged embedded actuation to fold low-profile sheets into 3D structures using materials such as shape memory polymers~\cite{nisser2016feedback,felton2014method} and shape memory alloys~\cite{firouzeh2015robogami} as well as using external activation via uniform heating~\cite{tolley2014self} and selective laser heating~\cite{nisser2021laserfactory}. However, these require dedicated machinery such as lasers, ovens or local joule-heating to activate material actuators. These actuators can also be challenging to use as they must be sufficiently low-profile to be embedded in the sheet, reducing their strength and often limiting candidates to material-based actuators such as shape memory materials which are challenging to control and often cannot be activated bidirectionally. They are typically redundant after initial use, and the embedding of these actuators results in a thickening and weighing down of the laminate itself, compromising both foldability and portability. In contrast, our string-based design is highly inexpensive, requires no dedicated machinery, can be actuated powerfully via external means, and can be removed following fabrication.

Methods for unfolding 3D polyhedral meshes into 2D sheets has been  studied across mathematics, physics, mechanical engineering and computer graphics \cite{takahashi2011optimized,demaine2005survey}. While prior work has explored pull-up nets to fold 3D geometries~\cite{meenan2008pull}, this was restricted to manual fabrication designs for the 5 platonic solids; in our work, we provide a tool to automatically compute these designs for all admissible geometries and introduce a digital fabrication pipeline to manufacture them. Given a 3D target structure, this process unfolds its 3D mesh into a planar 2D sheet populated with cutlines and throughholes. After laser-cutting the sheet and feeding thread through these throughholes to form a pull-up net, a user pulls on the thread to fold the sheet into the 3D structure. We introduce the fabrication process and build several prototypes demonstrating the method's ability to rapidly create a breadth of geometries suitable for low-fidelity prototyping across a wide range of applications. These include load-bearing stools, regular polyhedra and organic geometries such as the Stanford bunny, spanning a range of scales. We additionally develop a web-based simulation tool that translates 3D meshes into manufacturable 2D sheets, providing an existing set of 141 polyhedra for users to use as well as a feature allowing users to upload custom designs themselves.

This paper is structured as follows. We introduce related work in unpacking and folding between 2D sheets and 3D structures, and introduce the algorithmic framework to our approach. We demonstrate their use in our user interface which permits uploading 3D meshes and simulating the unfolding and folding stages. We highlight the digital fabrication pipeline for fabricating structures based on the outputs of the user interface, and showcase a variety of fabricated geometries that span both aesthetic and functional applications in the space of low-fidelity rapid prototyping. Finally, we highlight our method's limitations and point to future avenues of research.


In summary, our paper contributes:
\begin{itemize}
    \item An algorithm to unfold a 3D mesh into a planar 2D sheet populated with cutlines and throughholes compatible with our pull-up net technique.

    \item A web-based user interface, equipped with both a pre-existing set of 141 polyhedral meshes and a feature to accept custom 3D meshes, that outputs manufacturable 2D sheets for use on a laser cutter.
  

    \item A range of fabricated geometries that demonstrate our pull-up net method's ability to rapidly create both aesthetic and functional geometries spanning a range of applications in low-fidelity rapid prototyping.
    
\end{itemize}

\begin{figure}
  \includegraphics[width=\linewidth]{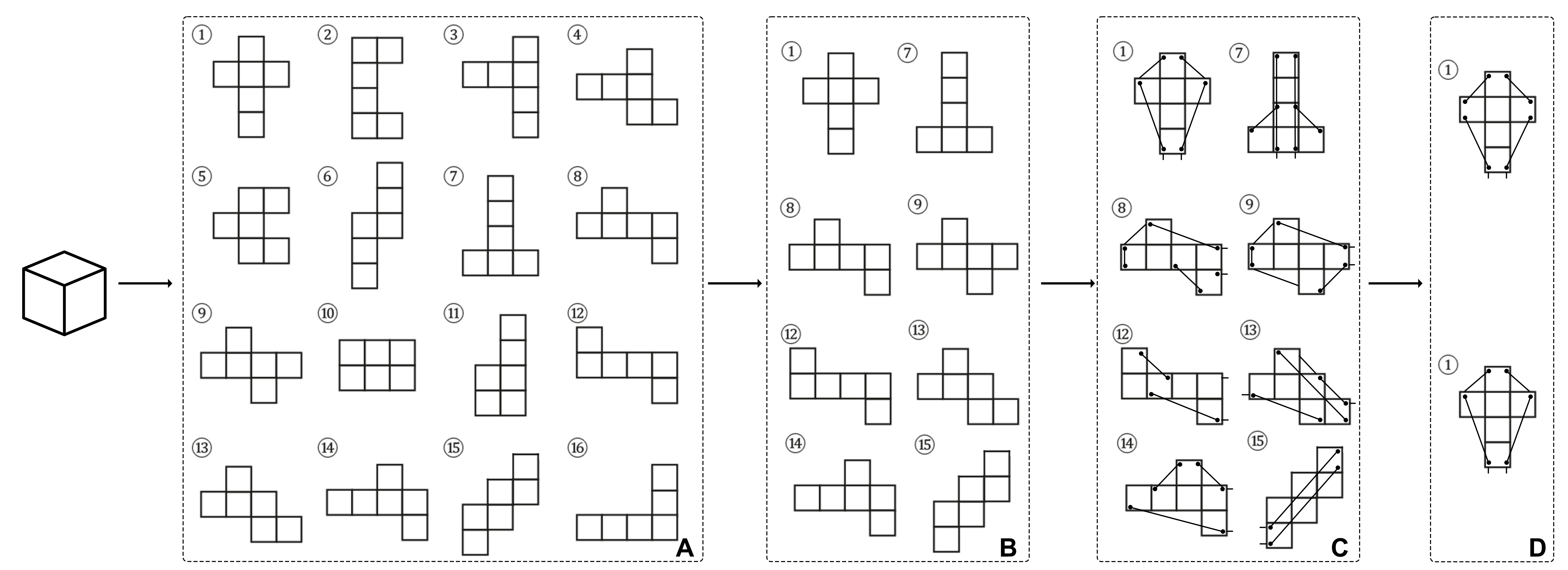}
  \caption{Unfolding a cube. (A) There are multiple potential ``nets'' for a cube but not all can be successfully pulled up to become a cube again. (B) Potential successful ways for folding up from a 2D net to a cube. (C) Location of the string to pull up 2D net back to the 3D cube (D) Even for the same 2D net, there can be many ways to route the string in order to to fold the 3D shape.) }
  \label{fig: }
\end{figure}

\section{Background and related work}

\subsection{Fabricating Folded 3D Structures using Laser Cutters}

A range of prior works, particularly in HCI, has explored ways to enable the fabrication of 3D structures by folding laser-cut 2D sheets. Flaticulation~\cite{10.1145/3526113.3545695} used a set of cut-in-place joint patterns to enable users to fold laser-cut sheets into target shapes via articulated angles. crdbrd~\cite{hildebrand2012crdbrd} used a laser cutter to fabricate mutually intersecting planar cut-outs from 3D shapes which can be assembled to form low fidelity 3D prototypes. LaserOrigami~\cite{mueller2013laserorigami} and LaserFactory~\cite{nisser2021laserfactory} fabricated 2.5D objects in a laser cutter by attenuating the laser power to heat an acrylic substrate until it becomes compliant enough to fold due to gravity. LaserStacker~\cite{umapathi2015laserstacker} similarly used attenuated laser power to weld sheets of acrylic together to form low-fidelity 3D prototypes. HingeCore~\cite{abdullah2022hingecore} used sandwiched materials laser-cut with finger joints to a partial depth in order to permit a single sheet to be assembled via folding without the need of glue or tabs. InfOrigami~\cite{tao2021inforigami} unfolds 3D meshes into laser-cuttable and color-coded 2D sheets which can be assembled by users with the help of glue.  

However, methods to date may be difficult to scale to high-resolution structures with hundreds or thousands of folds. Structures manufactured using methods that consolidate the folding procedures within the laser cutter platform can be constrained by occlusion between target hinges and the laser. Similarly, user-folded methods may become burdensome and unwieldy by requiring
users to manually fold each hinge individually. In this work, we propose laser-cut structures that can be folded in a single step, by actuating a single degree of freedom transmitted via a pull-up string.

\subsection{Algorithms for Polyhedra Unfolding}
\par
The problem of unfolding a polyhedron into a non-overlapping net has challenged researchers for decades, and virtually no guarantees exist for unfolding general polyhedra \cite{demaine2007geometric,demaine2005survey}. The computational effort needed to search all possible unfoldings suffers from combinatorial explosion as the size of the polyhedron increases. Even more challenging is the solving the full ``blooming'' problem, which seeks to avoid collisions between faces as the polyhedron is unfolded \cite{song2004motion,biedl2005can}. Nevertheless, for simple polyhedral meshes, and for meshes with small dihedral angles, heuristic methods offer promising results \cite{schlickenrieder1997nets,chen2013edge}. Although few guarantees exist for finding non-overlapping nets, existing heuristic algorithms have succeeded on thousands of polyhedra searched so far, including shapes with thousands of vertices and faces \cite{demaine2007geometric,schlickenrieder1997nets}. Certain guarantees exist for finding unfoldings, including restriction to convex polyhedra where faces may also be cut \cite{aronov1991nonoverlap}, but the general problem of unfolding along edges remains open. In this work, we introduce a set of heuristic algorithms for constructing an optimal net and string path for the unfolded sheet of a given 3D mesh.

\begin{figure}[H]
  \includegraphics[width= 0.8\linewidth]{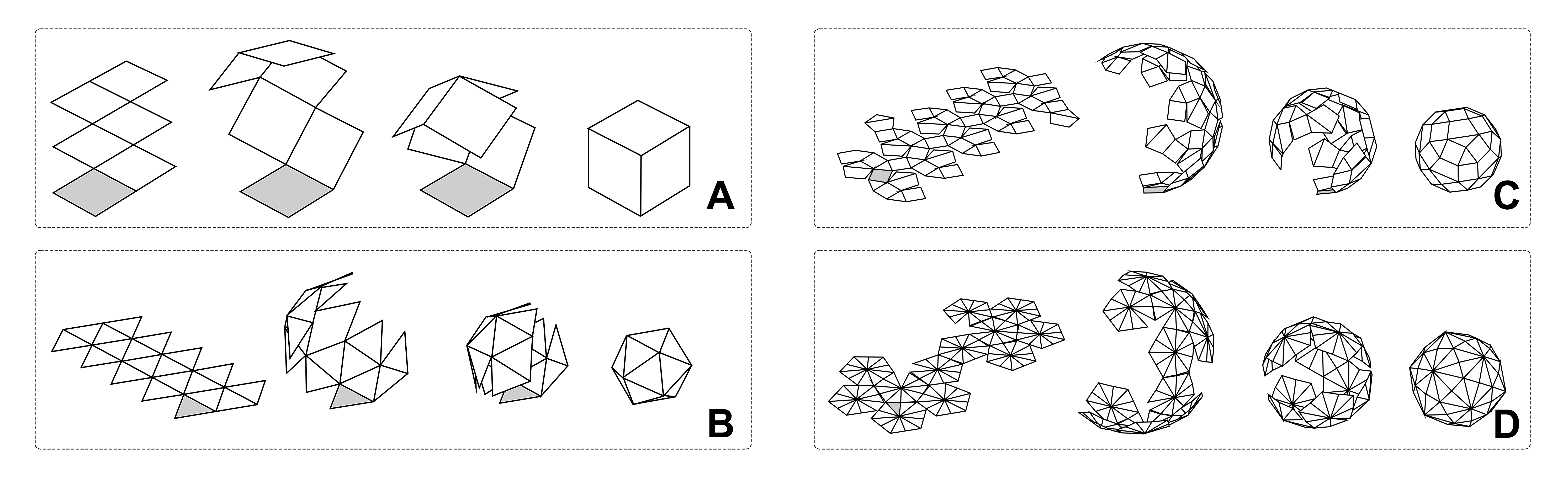}
  \caption{Examples of folding 3D polyhedra using pull-up nets: By convention, one surface (in grey) of the unfolding nets remains fixed during the folding procedure, helping stabilize the structure while users actuate the fold.}
  \label{fig: Examples}
\end{figure}

\subsection{Fabricating Folded Structures using Embedded Actuation}

Several actuation mechanisms in HCI, robotics, and art make use of actuators to connect disjoint faces and fold them along specific paths. Rivera et al.~\cite{rivera2017stretching} combined 3D-printing with textile-embedded, string-actuated mechanical arms to fold flat structures into 3D shapes. Self-shaping Curved Folding~\cite{tahouni2020self} 3D-printed hygroscopic materials in flat sheets that form curved 3D structures on exposure to moisture. Work on self-folding robots has led to the development of composite laminate materials, fabricated by bonding together layers of different materials that include structural layers, embedded electronics, actuators and flexure hinges, where actuators have ranged from shape memory polymers~\cite{nisser2016feedback} and shape memory alloys~\cite{firouzeh2015robogami} to pneumatically activated polymer “pouches”~\cite{niiyama2015pouch}. These laminates are composed entirely of thin layers with embedded actuators, whose 2D layouts are made using CAD (Computer Aided Design) software and bonded together, with the exception of discrete electronic components that are soldered on to circuit layers that are part of the composite. In work closely related to our own, Kilian et al~\cite{kilian2017string,meenan2008pull} computed and fabricated string actuation networks for a range of well known but limited crease patterns for folded surfaces.

Despite the many advances in using origami-inspired folding to acquire target structures and robot geometries, the embedded actuators have to date been difficult to control, and added significantly to the material thickness and complexity; in parallel, unfolding techniques for string-actuated structures have been shown for only few target shapes. In this paper we introduce a set of algorithms to fold a wide variety of 3D structures using pull-up nets: we contribute a software tool to automatically compute these designs for wide variety of admissible geometries, including 141 pre-computed designs, and introduce a digital fabrication pipeline to manufacture them.

\begin{figure}[H]
	\centering
	\includegraphics[width=0.7\linewidth]{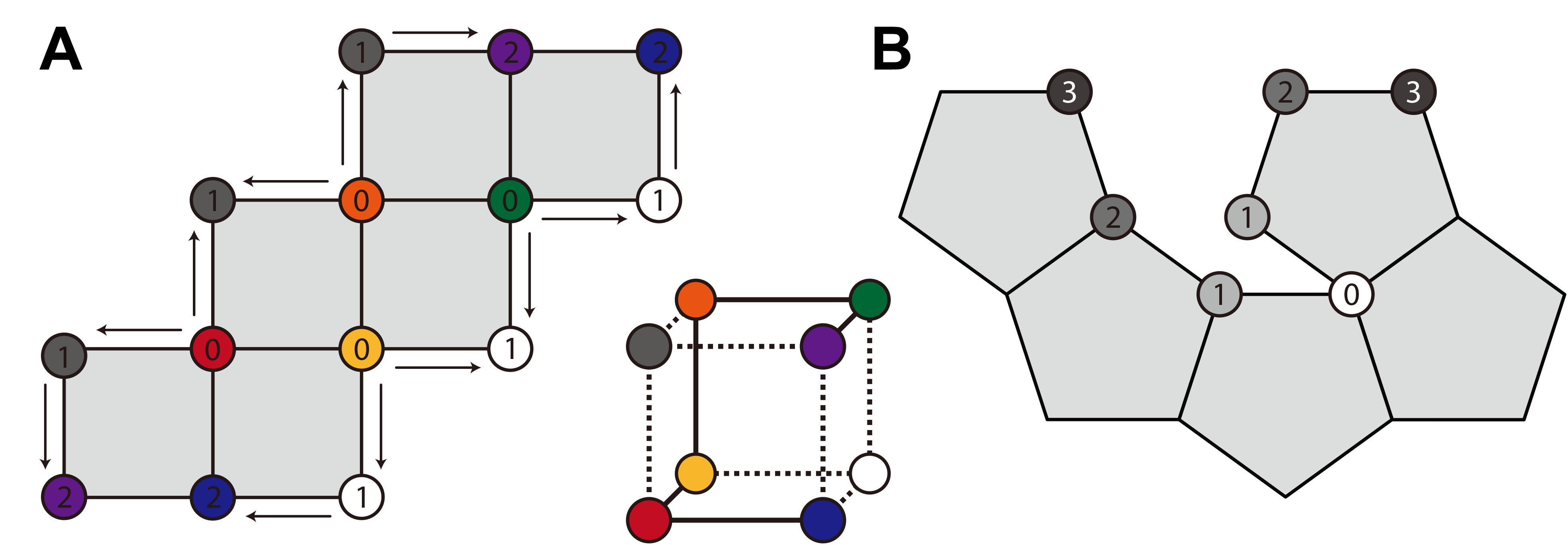}
	\caption{
        (A) Illustration of the algorithm to identify vertex sets that need to be joined by string. Color represents vertices of the net that are joined; numbers represent the rigidity depth of each vertex. Small arrows indicate the direction of the search from each vertex. Additionally, vertices of rigidity depth 1 do not need to be joined by string in order to assemble the final cube. In the 3D inset, cut edges are indicated by dotted lines.
        (B) The same concept illustrated with a portion of a dodecahedron net. The vertices of depth 1 and 2 do not necessarily have to join in order to determine a fully-joined geometric structure. Adding strings between all of these vertex pairs would require at least twice the amount of string, and would likely need to accommodate several sharp turns in its path, therefore increasing friction during assembly and complicating the assembly process. Depending on the assembled object's function and expected load, this may be a worthwhile tradeoff for extra rigidity in the final structure.
    }
	\label{fig:rigidity}
\end{figure}

\section{Algorithms}
This section introduces our algorithm for constructing an optimal net and string path for a given 3D mesh (Figure \ref{fig:rigidity}).

\textbf{Unfold the geometry} We begin by generating unfoldings of the mesh using a set of heuristic algorithms, including a steepest-edge cut tree, greatest-increase cut tree from Schlickenrieder~\cite{schlickenrieder1997nets} and a naive breadth-first unfolding from the largest face. If no non-overlapping nets are found, we randomly split the mesh in half (e.g., by cutting along edges that are nearest to a plane that passes through the center of mass) and repeat the search on each half.

\textbf{Join vertices} Next we use a breadth-first search to identify pairs of vertices that need to be joined (Figure \ref{fig:rigidity}): We identify vertices in the net that are already adjacent to $n$ connected faces, where the vertex is adjacent to $n$ faces in the 3D mesh. These vertices have a rigidity depth of 0. For each of these vertices: the vertex will be adjacent to two boundary edges in the net. The vertices at the other ends of these edges will need to be joined, and we sore these two vertices in a set and mark their rigidity depth as 1. Next, we join any sets with overlapping vertices. We finally iterate this joining procedure through the vertices of rigidity depth 1, until all vertices have been marked.

\textbf{Prune unwanted connections} We next remove unnecessarily joined vertices. The criteria for removing a vertex from a joined set are: (1) if removing the vertex from the set leaves only one vertex, then the other vertex must also be removed; (2) one of the vertices in the set must be connected by a single boundary edge to a joined vertex of greater rigidity depth; and (3) removing the vertex must leave each face with at least three joined vertices.

\textbf{Optimize string path} Using the final set of vertex sets to be joined, we lastly find the string path that passes through each pair of joined vertices which minimizes a combination of its turning angles and total length. If the mesh has been partitioned into multiple pieces, we optimize the string path on each piece separately.

\begin{figure} [H]
  \includegraphics[width= \linewidth]{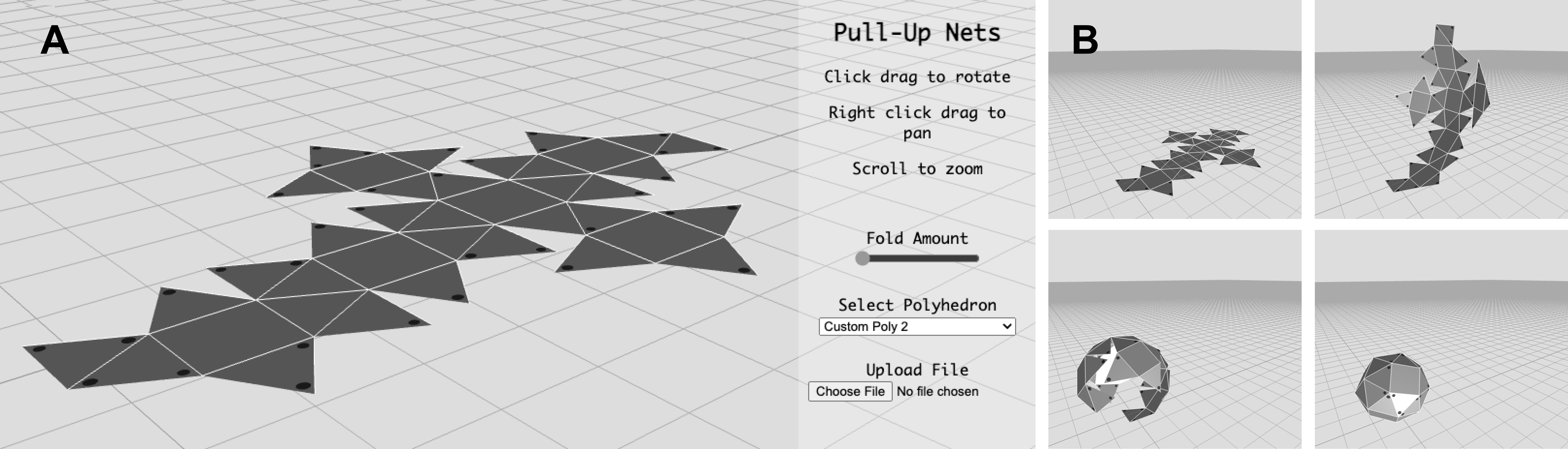}
  \caption{The web-based visualization tool.
  (A) The interface allows users to upload or select a geometry file, apply the folding algorithms to the geometry, and render the computed net with holes at each stage in its folding procedure.
  (B) Steps in the folding process can be visualized.}
  \label{fig:user_interface}
\end{figure}

\section{Web Tool}
We developed a web-based software tool based on our unfolding algorithm which allows users to upload their own custom 3D meshes from which to generate folded structures to fabricate. Users may upload either 3D mesh files in the OBJ file format, or existing nets in the SVG format; non-manifold meshes are rejected. In addition, an existing repertoire of 141 meshes that we successfully unfolded can be selected from a drop-down menu. The tool executes the algorithm on the chosen mesh and displays the mesh in the visualization window. Users can then render the folding procedure using a sliding toggle. 

The interface uses linear interpolation of edge folding angles to animate the assembly of the final shape. A ``base'' face is chosen arbitrarily to remain stationary. We use the Netlib polyhedra database~\cite{netlib} to avoid calculating unfoldings for many well-known simple polyhedra, although the Netlib unfoldings may only be used for convex polyhedra that do not allow faces to intersect. Animation and controls are made using the \verb|THREE.js| Javascript library.

\section{Fabricated Structures and Applications}

\begin{figure}[H]
  \includegraphics[width=0.5\linewidth]{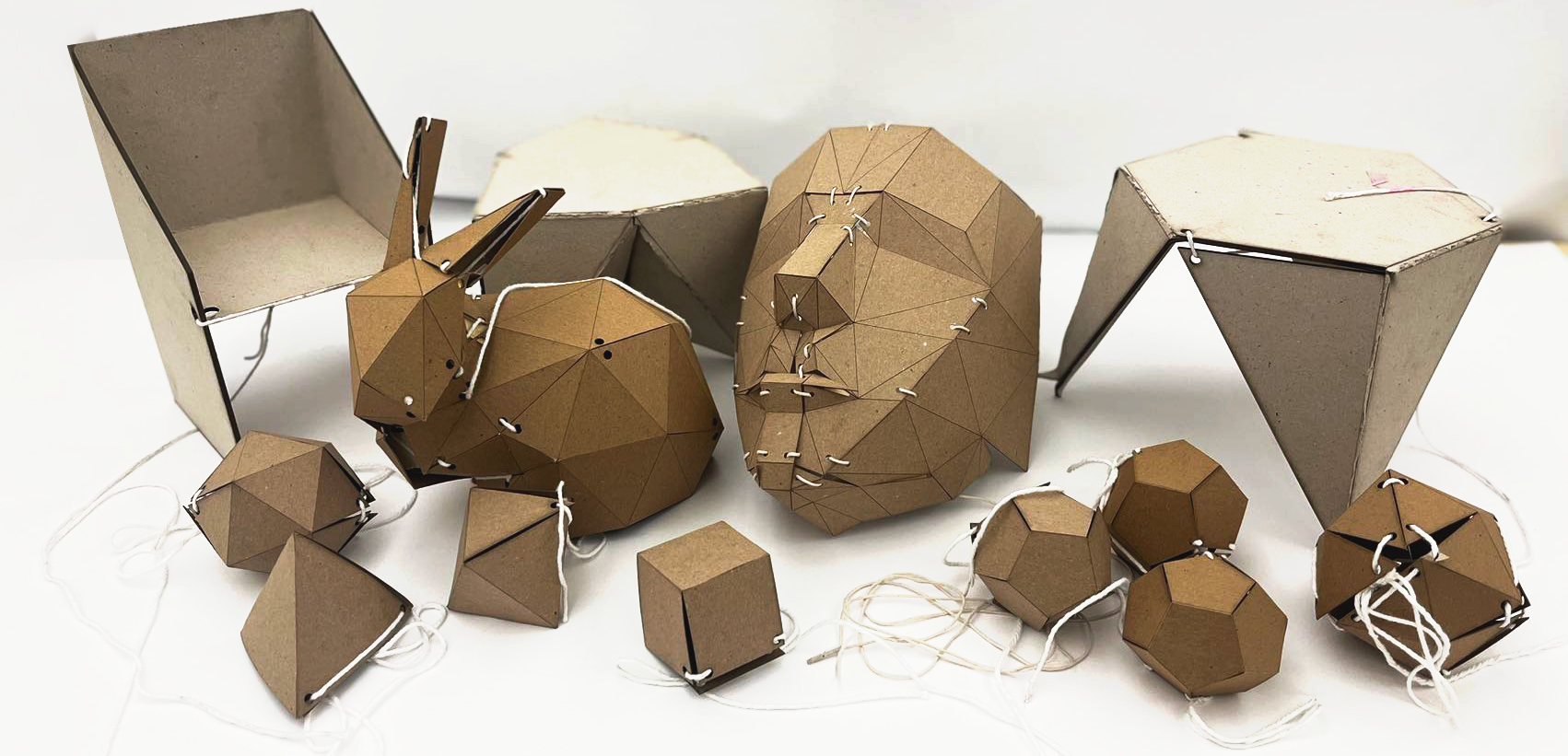}
  \caption{Structures fabricated using our pipeline range from regular polyhedra to organic shapes, and from load-bearing to aesthetic, across a range of scales, for applications that include VR, education and functional load-bearing contexts. }
  \label{fig: family}
\end{figure} 

\begin{figure}[H]
  \includegraphics[width=0.9\linewidth]{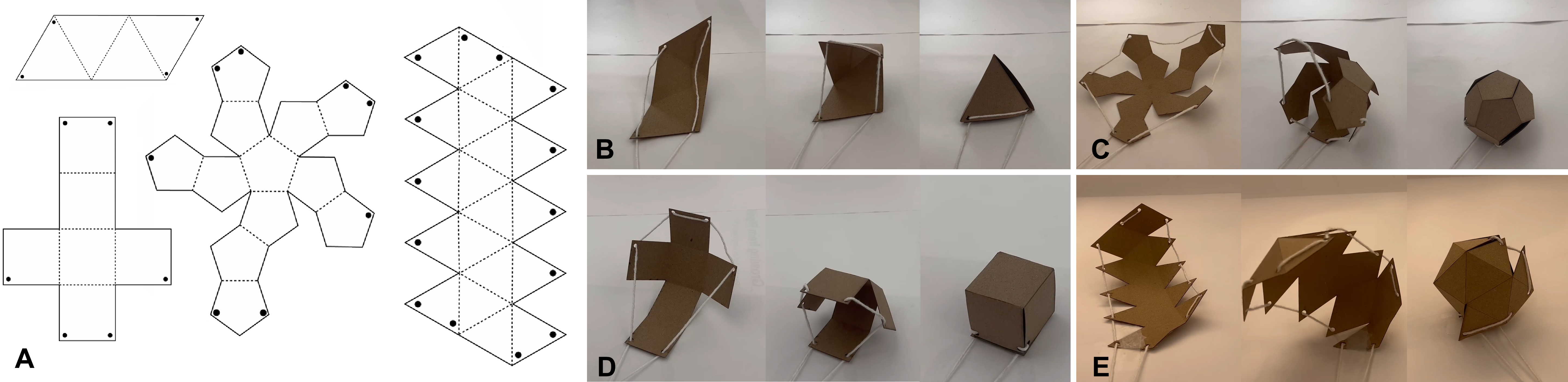}
  \caption{Fabricating polyhedra. (A) Unpacked layouts of four simple 3D solids. (B,C,D,E) Fabricated structures during folding.}
  \label{fig:fabricated-Polyhedra}
\end{figure}
\begin{figure}[H]
  \includegraphics[width=0.9\linewidth]{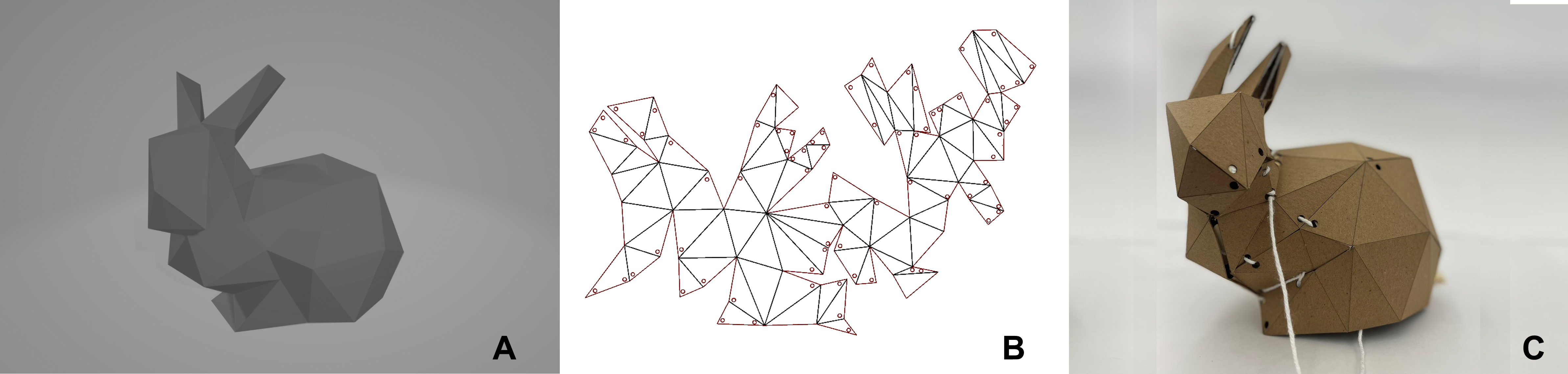}
  \caption{Fabricating organic shapes. (A) Rendering, (B) unpacking, and (C) fabrication of a Stanford bunny.}
  \label{fig:fabricated-Organic}
\end{figure}
\begin{figure}
  \includegraphics[width=\linewidth]{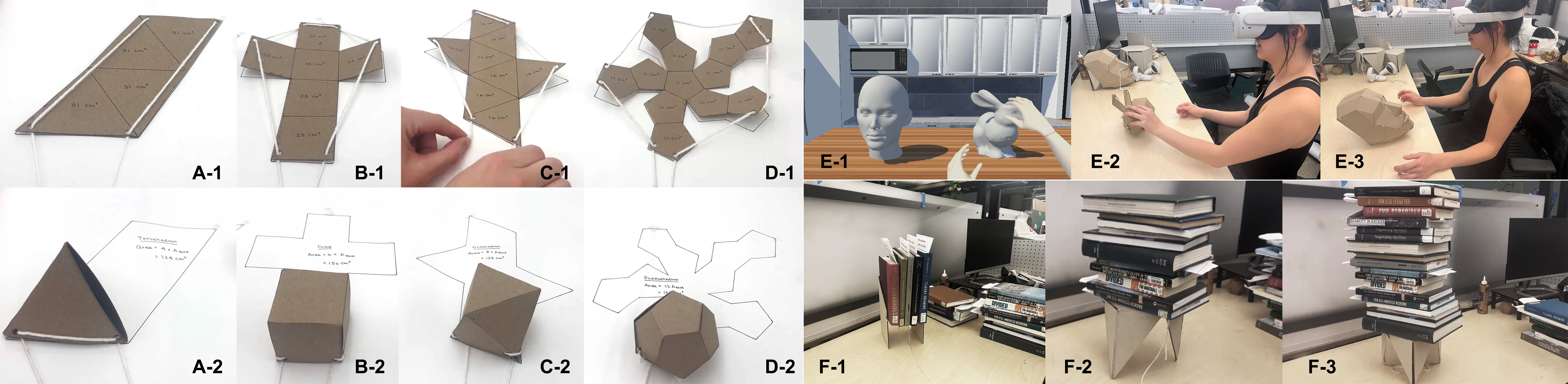}
  \caption{Applications include (A-D) Educational Toolkits to teach relationships between area and volume, (E) on-demand fabrication of low-resolution physical props for Virtual Reality, and (F) load-bearing applications including bookshelves or stools.}
  \label{fig:application}
\end{figure}

We fabricated a range of structures (Figure \ref{fig: family}) that are both aesthetic and load-bearing, including both traditional polyhedra (Figure \ref{fig:fabricated-Polyhedra}) and organic shapes (Figures \ref{fig:fabricated-Organic},\ref{fig: teaser}) across a range of scales to showcase the versatility of this technique in rapidly producing complex geometries. Possible applications include \textit{Educational Toolkits} (Figure \ref{fig:application}A-D), for example to teach children square-cube law relationships between surface area and volume, here with Platonic Solids. Visual dominance over some tactile cues~\cite{abtahi2018visuo} can be exploited to permit rapidly fabricating and folding \textit{low-resolution physical props for Virtual Reality} (VR) on-demand that are overlayed with high-resolution images (Figure \ref{fig:application}E). This manufacturing process is also suitable for \textit{load-bearing applications}, including bookshelves or small stools (Figure \ref{fig:application}F). These structures further benefit from their ability to be unfolded for compact storage or transport.

To build these structures, we used a laser cutter to cut the unfolded geometries produced by our web tool. We manually add cut holes where our algorithm selects them to appear, and then route string through these using our web tool rendering as a guide. We used 1/16'' chipboard and traditional knitting yarn to route through holes. Once routed, we pulled the yarn to fold the sheets into their target shapes.

\section{Limitations and Future Work}
\par
Having presented our digital fabrication pipeline, we discuss its limitations and avenues for future work.

\textbf{Algorithm} Each step in our algorithm may be computationally costly without optimization, and is ideally suited for smaller meshes with $\lesssim$ 100 elements. As noted, not all meshes may be unfolded into a single flat piece~\cite{bern1999ununfoldable,reitebuch2019discrete}, an issue exacerbated for meshes with excessively hyperbolic vertices (where all adjacent faces meet with angles summing to more than $2\pi$), common in geometries where convex polyhedral faces are extruded, or in periodic minimal surfaces. Although other assembly methods for such geometries exist \cite{overvelde2016three}, we do not yet consider them suitable for our method.

\textbf{String friction} Friction limits the maximum length of string used, the severity of turning angles as it passes through holes, and the number of holes present. Like a shoelace woven through holes at tight angles, friction may prevent tension at one end of the string from diffusing equally to all other parts. During assembly of polyhedra with many faces, the string may therefore require tension at several different points along its length to decrease frictional resistance.

\textbf{Web interface} The interface generates vertex locations for routing string, but does not yet generate ready fabrication files with holes at vertex locations, or indicate how to route string through each vertex, though this has been straightforward to interpret for our geometries produced so far. Further work will seek to automate this for fabrication on a laser cutter, and to implement basic mesh editing features for coarsening, triangulation, and refinement.

\textbf{Structure rigidity} The constraint of small-complexity solids is enforced physically as structural rigidity decreases as more cuts are introduced. Due to the large number of cuts needed for a complete unfolding (for a closed mesh, the cuts must reach every non-flat vertex), the complexity of some meshes may compromise its structure's stability, further descriptions of which we leave to future works. Additionally, there is a trade-off between reducing the number of vertices explicitly joined by the string path and reducing the rigidity of the final structure (Figure~\ref{fig:rigidity}). Another trade-off lies in the material thickness: thicker material exacerbates folding difficulty while thinner material is less stiff and tolerant to load-bearing. In future work we will seek to physically evaluate the load-bearing capacities of our structures.

\textbf{String path} A single 3D mesh can often be unfolded into different 2D surfaces (Figure \ref{fig: Examples}C). While we optimize string path by length and turning angles, we found that our algorithm's selected design is not always the easiest to pull up; more work is required to understand how to optimize for ease of fabrication.


\section{Conclusion}

In this paper, we introduced a method to rapidly create 3D geometries from 2D sheets using pull-up nets. We developed an algorithm run by our web-based software tool that allows users to upload 3D meshes and generate their unfolded geometries. We show how to fabricate these on a commercial laser cutter and route string through the faces which are pulled by a user to fold the sheet into its target 3D structure. This lets users rapidly create large 3D geometries using 2D fabrication machines using just a single actuated degree of freedom. We fabricated a variety of polyhedra and organic structures, highlighted applications, and laid out the work's current limitations and avenues for future work. We believe this work introduces a low-fidelity rapid prototyping option for 3D structures that is simple, inexpensive and rapid, obviating the need for significant design or fabrication expertise, and serves as a compelling alternative to more complicated material actuator-based folding paradigms. 

\bibliographystyle{ACM-Reference-Format}
\bibliography{sample-base}

\end{document}